\documentclass[preprint,showpacs,preprintnumbers,amsmath,amssymb]{revtex4}

\usepackage{graphicx}% Include figure files
\usepackage{epsfig}		
\usepackage{dcolumn}% Align table columns on decimal point
\usepackage{bm}% bold math

\def\0{\mbox{\tiny $0$}}
\def\1{\mbox{\tiny $1$}}
\def\2{\mbox{\tiny $2$}}
\def\3{\mbox{\tiny $3$}}
\def\4{\mbox{\tiny $4$}}
\def\5{\mbox{\tiny $5$}}
\def\6{\mbox{\tiny $6$}}
\def\7{\mbox{\tiny $7$}}
\def\8{\mbox{\tiny $8$}}
\def\9{\mbox{\tiny $9$}}

\def\n{\mbox{\tiny $n$}}
\def\k{\mbox{\tiny $k$}}

\def\e{\mbox{\tiny $e$}}

\def\f14{\mbox{\tiny $\frac{1}{4}$}}

\def\L{\mbox{\tiny $L$}}

\def\ii{\mbox{\tiny $i$}}
\def\l{\mbox{\tiny $l$}}

\def\P{\mbox{\tiny $P$}}

\def\j{\mbox{\tiny $j$}}
\def\mi{\mbox{\tiny $-$}}

\def\bb#1{\mbox{\footnotesize $(#1)$}}
\begin{document}

\title{Dirac neutrino mass from the beta decay end-point modified by the dynamics of a Lorentz-violating equation of motion.}

\author{A. E. Bernardini}
\email{alexeb@ifi.unicamp.br}
\affiliation{Instituto de F\'{\i}sica Gleb Wataghin, UNICAMP,
PO Box 6165, 13083-970, Campinas, SP, Brasil.}

\date{\today}% It is always \today, today,

\begin{abstract}
Using a generalized procedure for obtaining the dispersion relation and the equation of motion for a propagating fermionic particle, we examine previous claims for a lightlike preferred axis embedded in the framework of Lorentz-invariance violation with preserved algebra.
We show that, in a relatively high energy scale, the corresponding equation of motion is reduced to a conserving lepton number chiral equation previously predicted in the literature.
Otherwise, in a relatively low energy scale, the equation is reduced to the usual Dirac equation for a free propagating fermionic particle.
The new dynamics introduces some novel ingredients to the phenomenological analysis of the tritium beta decay.
In particular, a modified cross section expression and the correspondent phenomenology of the end-point are evaluated.
\end{abstract}

\pacs{03.30.+p, 12.15.Ff, 23.40.-s}
\keywords{Lorentz Violating Systems - Neutrino Mass - Beta Decay}
\date{\today}

\maketitle
Theoretical issues pointing to the possibility that many empirical successes of
special relativity do need not demand Lorentz invariance of the underlying theory have been widely discussed in several frameworks \cite{Ame01,Mag02,Kos06,AmeDSR,Col97,Bet06,DeG06}.
The characteristic scale of such theories is likely to be associated with the Planck's energy $E_{\P\l} \sim 10^{\1\9}$ GeV.
However, the current attainable energies are minuscule compared to this scale, so that experimental signals are expected to be heavily suppressed.
and no decisive evidence contradicting the exact Lorentz invariance has yet been experimentally detected.
In this scenario, Cohen and Glashow pursue a different approach to the possible failure of Lorentz symmetry denominated {\em very special relativity} (VSR) \cite{Gla06A,Gla06B}.
It is based on the hypothesis that the space-time symmetry group of nature is smaller than the Poincar\'{e} group, consisting of space-time translations and one of certain subgroups of the Lorentz group.
The formalism of VSR has been expanded for studying some peculiar aspects of neutrino physics with the VSR subgroup chosen to be the 4-parameter group SIM(2) \cite{Gla06B}.
Since neutrinos are now known to be massive, several mechanisms have been contrived to remedy the absence of neutrino mass in the Standard Model Lagrangian \cite{Bil03}.
The framework of VSR admits the unconventional possibility of neutrino masses that neither violate lepton number nor require additional sterile states \cite{Gla06B}.
In the {\em Dirac} picture, lepton number is conserved with neutrinos acquiring mass via Yukawa couplings to sterile SU(2)-singlet neutrinos \cite{Zub98,Kim93}.
In the {\em Majorana} picture, lepton number is violated and neutrino masses result from a seesaw mechanism involving heavy sterile states or via dimension-6 operators resulting from unspecified new interactions \cite{Gel79,Moh86}.
In spite of not being Lorentz invariant, the lepton number conserving neutrino masses are VSR invariant.
There is no guarantee that neutrino masses have a VSR origin, but if so their sizes may be an indication of the magnitude of Lorentz-violating effects in other sectors, for instance, as a suggestion to the examination of the existence of a preferred axis in the cosmic radiation anisotropy \cite{Mag05}.

In this Letter we report about an adequacy of the results of VSR into a Lorentz-invariance violation system reconstructed by means of modified conformal transformations acting on the Lorentz generators.
By considering such frameworks allowing for the existence of a preferred frame, we derive expressions for the spectrum and the end-point of the beta decay, which can be used as an experimental probe of the peculiar way in which neutrinos experience Lorentz invariance.
To reach this objective we combine each boost/rotation with an specific transformation from which we introduce a preferential direction with the aid of a light-like vector defined as $n_{\mu}$($\equiv(1,0,0,1)$), $n^{2}=0$.
The transformation has to be chosen as to bring an equation of motion which recovers the dynamics of the equation introduced in \cite{Gla06B},
\small\begin{equation}
\left(\gamma^{\mu}p_{\mu} - \frac{m_{\nu}^{\2}}{2}\frac{\gamma^{\mu}n_{\mu}}{p_{\lambda}n^{\lambda}}\right)(1-\gamma^{\5}) u\bb{p_{\nu}} = 0,
\label{pp01}
\end{equation}\normalsize
which admits the unconventional possibility of neutrino masses that neither violate lepton number nor require additional sterile states.
We search for convenient unitary transformations $U$ acting on the usual Lorentz generators in order to recover the equation of motion for a free propagating fermionic particle.
We expect that, in a relatively high energy scale, the corresponding equation will be reduced to the Glashow Eq.(\ref{pp01}), and, in a relatively low energy scale, it will be reduced to the usual Dirac equation for a free propagating fermionic particle.

Let us start with the definition of the momentum space $\mathcal{M}$ as the four-dimensional vector space consisting of momentum vectors $p_{\mu}$.
The ordinary Lorentz generators act as
\small\begin{equation}
L_{\mu\lambda} = p_{\mu}\partial_{\lambda} - p_{\lambda}\partial_{\mu}
\label{pp02}
\end{equation}\normalsize
where $\partial_{\mu} \equiv \partial/\partial p^{\mu}$, and we assume the Minkowski metric signature and that all generators are anti-Hermitian (also $\mu,\,\lambda = 0,\,1,\,2,\,3$, and $i,\,j,\,k = 1,\,2,\,3$ and the velocity of the light $c = 1$).
The ordinary Lorentz algebra is constructed in terms of the usual rotations $J^{\ii}\equiv \epsilon^{\ii\j\k}L_{\j\k}$ and boosts $K^{\ii} \equiv L_{\ii\0}$ as
\small\begin{equation}
[J^{\ii}, K^{\j}] = \epsilon^{\ii\j\k}K_{\k};~~~~[J^{\ii}, J^{\j}] = [K^{\ii}, K^{\j}] = \epsilon^{\ii\j\k}J_{\k}
\label{pp02A}
\end{equation}\normalsize
In order to introduce the nonlinear action that modifies the ordinary Lorentz generators but, however,
preserves the structure of the algebra, we suggest the following {\em ansatz} for a generalized transformation,
\small\begin{equation}
D \equiv (a\bb{y}\,p_{\mu} + b\bb{y}\,n_{\mu})\partial^{\mu}
\label{pp03}
\end{equation}\normalsize
which acts on the momentum space as
\small\begin{equation}
D \circ p_{\mu} \equiv a\bb{y}\,p_{\mu} + b\bb{y}\,n_{\mu}
\label{pp04}
\end{equation}\normalsize
where $y = p_{\mu} n^{\mu}$.
We assume the new action can be considered to be a nonstandard and nonlinear embedding of the Lorentz group into a modified conformal group which, as we shall notice in the following for the case of main interest, despite the modifications, satisfies precisely the ordinary Lorentz algebra (\ref{pp02A}).
To exponentiate the new action, we note that
\small\begin{equation}
k^{\ii} = U^{^{\mi 1}}\hspace{-0.35 cm}\bb{D}\, K^{\ii}\, U\bb{D} ~~\mbox{and}
~~ j^{\ii} = U^{^{\mi 1}}\hspace{-0.35 cm}\bb{D} \,J^{\ii} \,U\bb{D}
\label{pp06}
\end{equation}\normalsize
where the $y$-dependent transformation $U\bb{D}$ is given by $U\bb{D\bb{y}} \equiv{\exp[D\bb{y}]}$.
The nonlinear representation is then generated by $U\bb{D\bb{y}} \circ p_{\mu}$ and, despite not being unitary ($U\bb{D\bb{y}} \circ p_{\mu} \neq p_{\mu}$), it has to preserve the structure of the algebra.
Thus, when we assume
\small\begin{equation}
\left[[L_{\mu\lambda},\,D\bb{y}],\,D\bb{y}\right] = 0
\label{pp07}
\end{equation}\normalsize
we can reobtain a set of generators (in terms of $k_{\ii}$ and $j_{i}$) which satisfy the ordinary Lorentz algebra of (\ref{pp02A}).
At this point, in order to explicitly obtain the operator $D\bb{y}$ which satisfies the relation (\ref{pp07}), we firstly compute the commuting relation
\small\begin{equation}
[L_{\mu\lambda},\,D\bb{y}] = \kappa_{\mu\lambda}(a^{\prime}\bb{y}\,p_{\alpha} + b^{\prime}\bb{y}\,n_{\alpha})\partial^{\alpha}
+ b\bb{y} \, d_{\mu\lambda},
\label{pp08}
\end{equation}\normalsize
for which we have defined the parameters $\kappa_{\mu\lambda} = p_{\mu} n_{\lambda} - p_{\lambda} n_{\mu}$ and $d_{\mu\lambda} = n_{\lambda} \partial_{\mu} - n_{\mu} \partial_{\lambda}$.
From the above definitions we obtain the useful relations
\small\begin{equation}
D \,\kappa_{\mu\lambda} = a\bb{y} \kappa_{\mu\lambda},~~
d_{\mu\lambda} \,D = a\bb{y} d_{\mu\lambda}~~ \mbox{and}~~
D \,d_{\mu\lambda} = 0,
\label{pp10}
\end{equation}\normalsize
which are essential in computing $[[L_{\mu\lambda},\,D\bb{y}],\,D\bb{y}]$.
The first part of the r.h.s. of the Eq.(\ref{pp08}) then leads to the commuting relation
\small\begin{eqnarray}
\lefteqn{[\kappa_{\mu\lambda}(a^{\prime}\bb{y}\,p_{\lambda} + b^{\prime}\bb{y}\,n_{\lambda})\partial_{\lambda},\,D\bb{y}] =}\nonumber\\
&& \kappa_{\mu\lambda}\left\{\left[y (a^{\prime \2}\bb{y} - a\bb{y}\, a^{\prime \prime}\bb{y}) - a\bb{y}\,a^{\prime}\bb{y}\right] p^{\lambda}\partial_{\lambda}\right.\nonumber\\
&&\left.~~~~+ \left[a^{\prime}\bb{y}(y\, b^{\prime}\bb{y} - b\bb{y}) -y\, a\bb{y}\,b^{\prime \prime}\bb{y}\right] n^{\lambda}\partial_{\lambda}\right\},
\label{pp11}
\end{eqnarray}\normalsize
and the second part gives
\small\begin{equation}
[b\bb{y} \, d_{\mu\lambda},\,D\bb{y}] = a\bb{y} (b\bb{y} - y \,b^{\prime}\bb{y}) d_{\mu\lambda}.
\label{pp12}
\end{equation}\normalsize
In order to satisfy the condition for preserving the Lorentz algebra (\ref{pp07}), the $y$-dependent coefficients can be obtained by evaluating the coupled ordinary differential equations:
\small\begin{eqnarray}
y (a^{\prime \2}\bb{y} - a\bb{y}\, a^{\prime \prime}\bb{y}) - a\bb{y}\,a^{\prime}\bb{y} = 0 &&~~~~(a.1)\nonumber\\
a^{\prime}\bb{y}(y\, b^{\prime}\bb{y} - b\bb{y}) -y\, a\bb{y}\,b^{\prime \prime}\bb{y}  = 0 &&~~~~(a.2)\nonumber\\
a\bb{y} (b\bb{y} - y \,b^{\prime}\bb{y})                                                = 0 &&~~~~(a.3)\nonumber
\end{eqnarray}\normalsize
for which we have two types of solutions:

\paragraph*{Type-I}  $a\bb{y}=0$ and $\forall ~b\bb{y}$.

\paragraph*{Type-II}  $a\bb{y}= A y^{\n}, n\in \mathcal{R}$ and $b\bb{y}= B y$ where $A$ and $B$
are constants with respective dimensions given by $[[A]]\equiv m^{\mi\n}$ and $[[B]]\equiv m^{\mi \1}$.

However, for Type-II solutions, it is difficult (and sometimes impossible) to constraint $M^{^{\2}}\bb{\alpha y}$ to
preserve the standard dispersion relation $p^{\2}= m^{\2}$.
This objection makes the above solutions not so interesting as the Type-I solution.

For a Type-I solution where $a\bb{y} = 0$ and
$b\bb{y} = - \alpha \,m^{\2}/(1 + 2 \alpha y)$,
we can easily verify that
\small\begin{eqnarray}
\lefteqn{D\bb{y} \equiv -\frac{\alpha}{1+ 2 \alpha y}\,n_{\mu}\partial^{\mu}
\Rightarrow D\bb{y}\circ p_{\mu}\equiv -\frac{\alpha\,m^{\2}}{1+ 2 \alpha y}\,n_{\mu}}  \nonumber\\
&&~~~~~~~~\Rightarrow U\bb{D\bb{y}}\circ p_{\mu}\equiv p_{\mu(\alpha)} = p_{\mu} - \frac{\alpha\,m^{\2}}{1+ 2 \alpha y}\,n_{\mu}.
\label{pp16}
\end{eqnarray}\normalsize
In spite of preserving the Lorentz algebra \cite{Ber07},
these transformations clearly do not preserve the usual quadratic invariant in the momentum space.
But there is a modified invariant $||U\bb{D\bb{y}}\circ p_{\mu}||^{\2} = M^{^{\2}}\hspace{-0.15cm}\bb{\alpha}$ which leads to
the following dispersion relation,
\small\begin{eqnarray}
||U\bb{D\bb{y}}\circ p_{\mu}||^{\2} = p_{(\alpha)}^{\2} = p^{\2} - \frac{2\, y\,m^{\2}\,\alpha}{1+ 2 \alpha y} =M^{^{\2}}\hspace{-0.15cm}\bb{\alpha}
\label{pp17}
\end{eqnarray}\normalsize
Imposing the constraint $p^{\2}= m^{\2}$, which is also required by the VSR theory, we have the Casimir invariant
\small\begin{equation}
M^{^{\2}}\hspace{-0.15cm}\bb{\alpha} = \frac{m^{\2}}{1+ 2 \alpha y}.
\label{pp18}
\end{equation}\normalsize\normalsize
for which the $U$-invariance can be easily verified when we apply the transformation $U\bb{D\bb{y}}$.
The fact that the algebra of the symmetry group remains the same suggests that perhaps the standard spin connection formulation of relativity is still valid.
In this sense, the above dispersion relation can also be obtained from the dynamic equation for a fermionic particle,
\small\begin{eqnarray}
\lefteqn{\left[\gamma^{\mu}\left(U\bb{D\bb{y}}\circ p_{\mu}\right) - M\bb{\alpha}\right]u_{_{\L}}\bb{p_{\nu(\alpha)}} = 0} \nonumber\\
&&\Rightarrow\left(\gamma^{\mu}p_{\mu} - \frac{m^{\2} \alpha}{1 + 2\,\alpha\,y }\gamma^{\mu}n_{\mu} - M\bb{\alpha} \right)\,u_{_{\L}}\bb{p_{\nu(\alpha)}} = 0.
\label{pp19}
\end{eqnarray}\normalsize
Alternatively, as pointed out in \cite{Mag03}, for a comparative purpose to all these classes of Lorentz-violating models,
dispersion relations may be derived from calculations in a theory such as loop quantum gravity \cite{Gam99}.

By setting $[[m]]^{\mi\1}$ values to $\alpha$, for instance, $\alpha = \pm 1/m,\,\pm m/\varepsilon^{\2}_{\P\l}$
(where $\varepsilon_{\P\l}$ is the Planck energy),
we are able to analyze the low and the high energy limits.
In the high energy limit where $\alpha y >> 1$, the Eq.~(\ref{pp19}) is reduced to
\small\begin{equation}
\left(\gamma^{\mu}p_{\mu} - \frac{m^{\2}}{2\,y }\gamma^{\mu}n_{\mu} - M\bb{\alpha}\right)u_{_{\L}}\bb{p_{\nu(VSR)}} = 0.
\label{pp19A}
\end{equation}\normalsize
and since $M^{^{\2}}\hspace{-0.15cm}\bb{\alpha}\approx \frac{m^{\2}}{2\,|\alpha\,y|} << m^{\2}$,
in spite of not being necessary, we can eliminate the dependence on $\alpha$ since
the $M\bb{\alpha}$ term becomes irrelevant as we can observe when we take the quadratic form of the above equation.
Thus we recover the {\em VSR Cohen-Glashow} equation (\ref{pp01})\cite{Gla06B}
and its corresponding dispersion relation, as we have proposed from the initial part of this letter.
In the low energy limit where $\alpha y << 1$ ($p_{\nu(\alpha)}\rightarrow p_{\nu}$), the Eq.~(\ref{pp19}) is reduced to
\small\begin{equation}
\left(\gamma^{\mu}p_{\mu} - m^{\2}\,\alpha \gamma^{\mu}n_{\mu} - m \right)u_{_{\L}}\bb{p_{\nu}} = 0,
\label{pp19B}
\end{equation}\normalsize
whose the quadratic form is
\small\begin{equation}
\left(p^{\2} - 2 m^{\2}\,\alpha y - m^{\2}\right)u_{_{\L}}\bb{p_{\nu}} \approx  \left(p^{\2} -  m^{\2}\right)u_{_{\L}}\bb{p_{\nu}} = 0,
\label{pp19C}
\end{equation}\normalsize
i.e. when $\alpha y << 1$ the effective contribution from the second term of Eq.~(\ref{pp19B}) is minimal and the equation can be reduced to the usual (low energy limit) {\em Dirac} equation for a free propagating particle.
Such an important result could also be reproduced in a more direct way if we initially assumed
a natural energy scale where $m \alpha << 1$, for instance, when $\alpha =  m/\varepsilon^{\2}_{\P\l}$.
Since $M$ is reduced to $m$, the Eq.~(\ref{pp19B}) is immediately reduced to the {\em Dirac} equation.

\begin{figure}
\vspace{-0.6 cm}
\centerline{\psfig{file=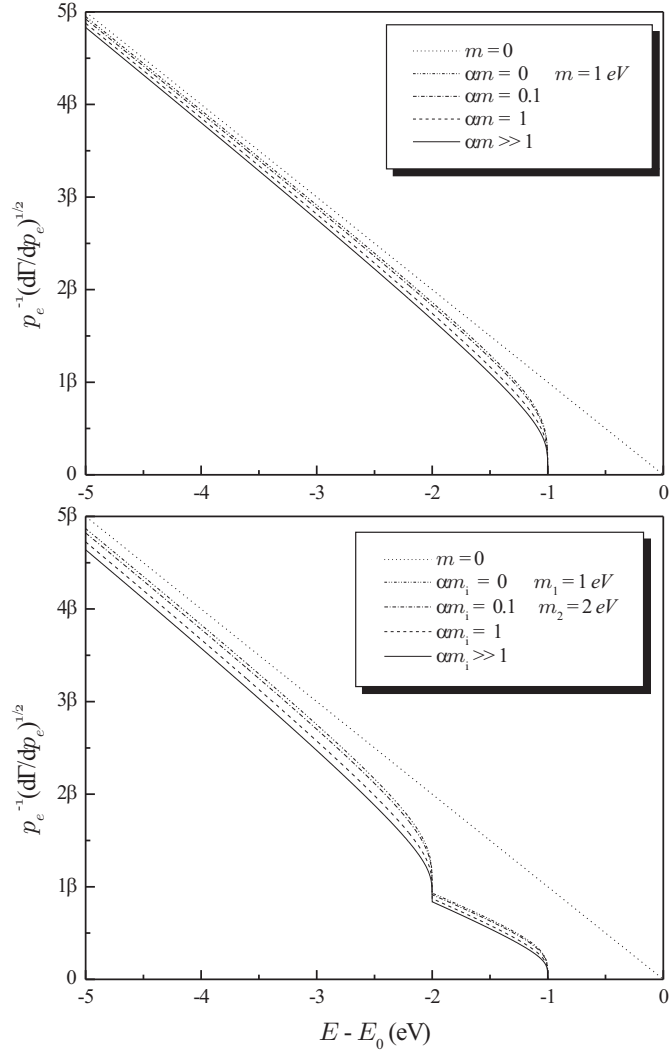,width=10.0 cm}}
\vspace{-1 cm}
\caption{
The Kurie plot rate $p_{\e}^{\mi\1}(\mbox{d} \Gamma/ \mbox{d}p_{\e})$ as a function of the neutrino energy ($E - E_{\0}$) near the end-point of the tritium beta decay spectrum for $m_{\nu} = 1\, $eV in the first plot,
and for $m_{\nu \1} = 1\, $eV and $m_{\nu \2} = 2\, $eV in the second plot.
The dotted line is for a massless neutrino.
The parameter $\alpha$ varies continuously from zero (the {\em low energy} or Dirac equation limit) to infinity (the {\em high energy} or the VSR limit).
Here $\beta$ is defined as $G/(2 \pi^{\3})^{\1/\2}$
\label{fig1}}
\vspace{-0.4 cm}
\end{figure}

\begin{figure}
\vspace{-0.5 cm}
\centerline{\psfig{file=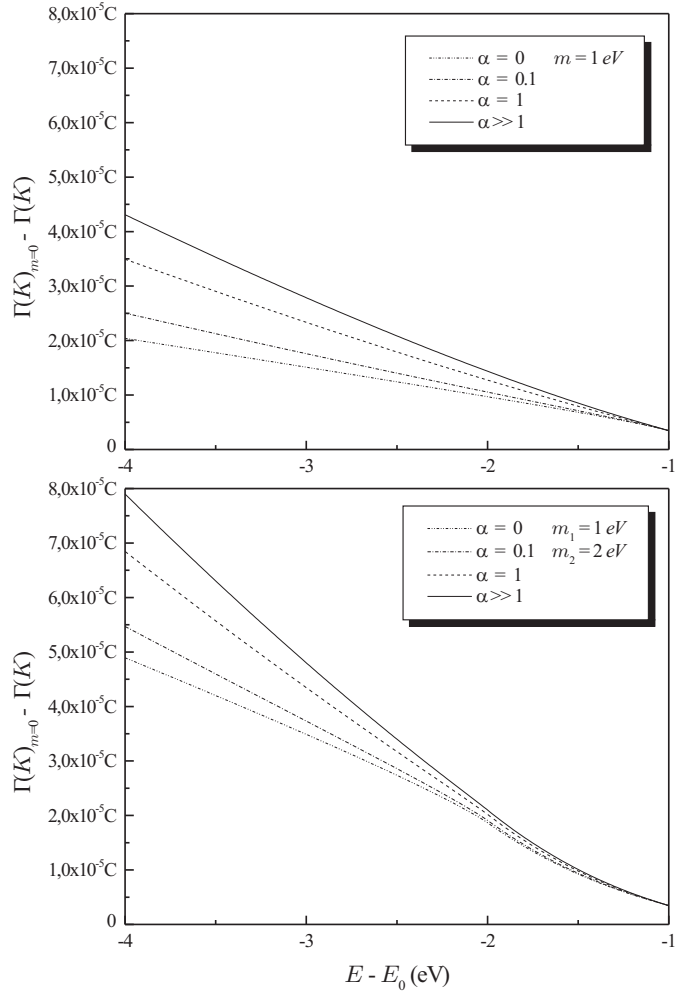,width=10 cm}}
\vspace{-1 cm}
\caption{
The integrated end-point difference $\Gamma_{m_{\nu}=\0}\bb{K} - \Gamma\bb{K}$ as a function of the neutrino energy ($- E_{\nu} = K - K_{max}= E - E_{\0}$) near the end-point of the tritium beta decay spectrum for $m_{\nu} = 1\, $eV in the first plot,
and for $m_{\nu \1} = 1\, $eV and $m_{\nu \2} = 2\, $eV in the second plot.
Again we vary the parameter $\alpha$ from zero to infinity
in order to obtain the {\em fine-tuning} correction to integrated energy spectrum from the end-point downward.
Here $C$ is defined as $G^{\2}/2 \pi^{\3}$
\label{fig2}}
\vspace{-0.5 cm}
\end{figure}

Once we have established the novel dynamics, the measurement of $\beta$-spectrum in the end-point region in tritium $\beta$-decay
is the classical method of direct determination of neutrino mass \cite{FerPer}.
Since the dispersion relations is maintained, the differential of the decay rate for the $d \rightarrow u\,e^{\mi}\,\bar{\nu}_{\e}$ transition is related to the decay amplitude by \cite{Hal84}
\small\begin{equation}
\mbox{d}\Gamma = G^{\2} \sum_{spins}\left|\bar{u}\bb{p_{\e(\alpha)}}\gamma^{\0} (1-\gamma_{\5})\upsilon\bb{p_{\nu(\alpha)}}\right|^{\2}
\frac{\mbox{d}^{\3}p_{\e}}{(2\pi)^{\3} E_{\e}}
\frac{\mbox{d}^{\3}p_{\nu}}{(2\pi)^{\3} E_{\nu}}
2\pi \delta\bb{E_{\0} - E_{\e} - E_{\nu}}
\label{pp30}
\end{equation}\normalsize
where $E_{\0}$ is the energy released to the lepton pair and $G$ is the Fermi constant.
For the novel dynamics, the phase space restriction is augmented by a change in the relevant matrix.
We see that the weak leptonic charged current $J^{\mu}$ must be modified to ensure its conservation,
\small\begin{equation}
J^{\mu} = \bar{u}\bb{p_{\e}}\left[\gamma^{\mu} +
\frac{\alpha^{\2}\,m_{\nu}^{\2}}{2}\frac{n^{\mu}(n^{\lambda} \gamma_{\lambda})}{(1 + 2 \alpha \, p_{\e}^{\lambda}n_{\lambda})(1 + 2 \alpha \, p_{\nu}^{\lambda}n_{\lambda})}\right]\upsilon\bb{p_{\nu}}
\label{pp31}
\end{equation}\normalsize
The Lorentz violating $\alpha$-dependent term leads to an entirely negligible effect near the end-point.
It yields a maximal correction of order $m_{\nu}/m_{\e}$ (when $\alpha \, m >>1$) which decreases as the parameter $\alpha\,m$ diminishes.
The first term, although not modified in form, differs from the standard current since the square matrix element $\upsilon\bb{p_{(\alpha)}}\bar{\upsilon}\bb{p_{(\alpha)}}$
is modified by
\small\begin{equation}
\upsilon\bb{p_{(\alpha)}}\bar{\upsilon}\bb{p_{(\alpha)}} = \frac{1-\gamma_{\5}}{2} p^{\mu}_{\nu(\alpha)}\gamma_{\mu}
= \left[p_{\nu}^{\mu} \gamma_{\mu} - \frac{\alpha\,m_{\nu}^{\2}}{2}\frac{n^{\mu} \gamma_{\mu}}{(1 + 2 \alpha \, p_{\nu}^{\lambda}n_{\lambda})}\right]
\label{pp32}
\end{equation}\normalsize
contrast to the standard expression.
Now the summation over the spins in the Eq.~(\ref{pp30}) can be performed in order to give
\small\begin{equation}
\sum_{spins}\left|\bar{u}\bb{p_{\e}}\gamma^{\0} (1\mi\gamma_{\5})\upsilon\bb{p_{\nu}}\right|^{\2} =
8\left[E_{\e}\left(E_{\nu} - \frac{n_{\0}\, \alpha \, m_{\nu}^{\2}}{(1 + 2 \alpha \, p_{\nu}^{\lambda}n_{\lambda})}\right)+
p_{\e} \left(p_{\nu}\cos(\theta_{\e}\mi\theta_{\nu}) - \frac{\cos(\theta_{\e})\, \alpha \, m_{\nu}^{\2}}{(1 + 2 \alpha \, p_{\nu}^{\lambda}n_{\lambda})}\right)\right]
\label{pp33}
\end{equation}\normalsize
where $\theta_{\e(\nu)}$ comes from the scalar product $\mathbf{p}_{\e(\nu)}\cdot\mathbf{n} = p^{\e(\nu)} \cos(\theta_{\e(\nu)})$.
Substituting  (\ref{pp33}) into (\ref{pp30}), the transition rates becomes
\small\begin{eqnarray}
\mbox{d}\Gamma &=& \frac{16 G^{\2}}{(2 \pi)^{\5}} \left[E_{\e}\left(E_{\nu} - \frac{n_{\0}\, \alpha \, m_{\nu}^{\2}}{(1 + 2 \alpha \, p_{\nu}^{\lambda}n_{\lambda})}\right)+
p_{\e} \left(p_{\nu}\cos(\theta_{\e}\mi\theta_{\nu}) - \frac{\cos(\theta_{\e})\, \alpha \, m_{\nu}^{\2}}{(1 + 2 \alpha \, p_{\nu}^{\lambda}n_{\lambda})}\right)\right]\nonumber\\
&&~~~~\times\left[\left(2\pi (\mbox{d}\cos(\theta_{\e}))p^{\e\2}\mbox{d}p^{\e}\right)\left(2\pi (\mbox{d}\cos(\theta_{\nu}))p_{\nu}^{\2}\mbox{d}p_{\nu}\right)\right]
\delta\bb{E_{\0} - E_{\e} - E_{\nu}}
\label{pp34}
\end{eqnarray}\normalsize
where we have rewritten $\mbox{d}^{\3}p_{\e(\nu)}/(2 E_{\e(\nu)})$ in terms of the spherical coordinates.
By performing the angular integrations and the $E_{\nu}\bb{p_{\nu}}$ integration, after some mathematical manipulations, we obtain
\small\begin{equation}
\frac{\mbox{d}\Gamma}{\mbox{d}p} = C\,  p^{\2}\, \left\{E_{\nu}\, p_{\nu}
- \frac{m_{\nu}^{\2}}{2}\ln{\left[\frac{1 + 2 \alpha (E_{\nu} + p_{\nu})}{1 + 2 \alpha (E_{\nu} - p_{\nu})}\right]}\right\}
\label{pp35}
\end{equation}\normalsize
where we have suppressed the electron index $e$ from the kinematical variables.
In the above equation, $E_{\nu} = E_{\0} - E = (K_{max} + m_{\e}) - (K + m_{\e})$, $p_{\nu}  = \sqrt{(K_{max} - K)^{\2} - m_{\nu}^{\2}}$ and the constant $C$ is defined as $G^{\2}/2 \pi^{\3}$.
The Kurie plot rate $p_{\e}^{\mi\1}(\mbox{d} \Gamma/ \mbox{d}p_{\e})$ as a function of the neutrino energy ($E - E_{\0}$) near the end-point of the tritium beta decay spectrum ($K_{\max} = 18.6\, $keV) for $m_{\nu} = 1\, $eV is shown in the first plot of the Fig.\ref{fig1}.
Since the final state neutrinos are not detected in the tritium $\beta$-decays experiments, for the electron spectrum, we get the incoherent sum
\small\begin{equation}
\frac{\mbox{d}\Gamma}{\mbox{d}p} = \sum_{\j=1}^{\2}|U_{\e\j}|^{\2} \frac{\mbox{d}\Gamma \bb{m_{\nu\j}}}{\mbox{d}p}
\label{pp36}
\end{equation}\normalsize
for a superposition of two neutrino mass eigenstates with $m_{\1} = 1\,$eV and $m_{\1} = 2\,$eV\footnote{Obviously these values can be changed in order to fit the experimental data.},
with mixing angle equal to $\pi/4$, in the second plot of the same figure.
It is important to notice that, in a realistic phenomenological analysis, we can vary continuously the parameter $\alpha$ from $0$ to $\infty$ (VSR dynamics limit with $\alpha \, m >> 1$),
as well as add the necessary number of neutrino species in (\ref{pp36}).
In fact, the knowledge of neutrino mass spectrum is decisive for the understanding of the origin of neutrino masses and mixing.
If in the KATRIN \cite{Katrin} experiment, which is under preparation at present, a positive effect due to the neutrino
mass will be observed, we will have $m_{\nu(\beta)} \approx m_{\nu_{\1,\2,\3}}$\footnote{If the effect of nonzero neutrino mass will not be observed, it will be crucial to improve the
sensitivity of the $\beta$-decay experiments.}.
The KATRIN experiment and its predecessors measured the integrate energy spectrum from the end-point downward which is proportional to
\small\begin{equation}
\Gamma\bb{K} = \int_{K}^{K_{max} - m_{\nu}}{\frac{\mbox{d}\Gamma}{\mbox{d}K}\mbox{d}K}
\label{pp37}
\end{equation}\normalsize
where $K(E_{\nu} = E - E_{\0})$ is implicitly defined above as the the electron kinetic energy $K = E - m_{\e} = K_{\max} + E_{\nu}$.
By observing that $K\,\mbox{d}K = E\,\mbox{d}E = p\,\mbox{d}p$, we can numerically evaluate the above integral in order
to obtain the Fig.\ref{fig2}.
The effect of neutrino mass is conveniently expressed as the difference
from the massless case in terms of $\Gamma_{m_{\nu}=\0}\bb{K} - \Gamma\bb{K}$ as a function of the neutrino energy ($E - E_{\0}$).
By knowing the experimental inputs, we could fit the results for the best values of $\alpha$ and $m$ and compare with the standard result ($\alpha = 0$).

To summarize, we have examined previous theoretical claims for a preferred axis at $n_{\mu}$($\equiv(1,0,0,1)$), $n^{2}=0$ in the framework of Lorentz invariance violation by generalizing the procedure for obtaining the equation of motion for a propagating fermionic particle.
We have shown that, in a relatively high energy scale, the corresponding equation of motion is reduced to a conserving lepton number chiral equation previously predicted in the literature \cite{Gla06B}, and, in a relatively low energy scale, it is reduced to the usual Dirac equation for a free propagating fermionic particle.
For fermionic particles with the dynamics described by the Eq.~(\ref{pp10}), the phase space kinematic restriction is modified by a change in the relevant matrix element.
In order to incorporate some discrete spatial and causal structures at the Planck energy scale, the action which leads to Lorentz invariance with an invariant energy scale ($\varepsilon_{\P\l}$ or $m^{\2}/\varepsilon_{\P\l}$) can be taken into account simultaneously with the action here proposed.
Ideally, the formalism we are discussing, in the sense analogous to that of the VSR, can be used to compare experiment and theory, as well as to extrapolate between predictions of different experimental measurements.
For this aim, we have quantified the eventual modifications to which could be observed in the next generation of tritium beta decay end-point experiments.
From the fit of the experimental data, it should be interesting obtaining an explicit relation between our free parameter $\alpha$ and the Planck energy.
Moreover, the formal construction presented in this letter can be used for parameterizing
two other possible phenomenologically observable modifications to neutrino physics:
(i) eventual modifications to the predictions to neutrinoless double beta decay \cite{Ver02}, and (ii) small modifications to the oscillation picture due to
Lorentz-violating interactions that couple with the active neutrinos and eventually allow the complete explanation of neutrino data.
With the compelling evidence for massive neutrinos from recent oscillation experiments, one of the most fundamental tasks of particle physics over the next years will be the determination of the absolute mass scale of neutrinos.
Beside of being part of a phenomenology of quantum gravity effects, as opposed to directly having a fundamental significance, this formalism may eventually contribute for determining the absolute value of neutrino masses, which will have crucial implications for cosmology, astrophysics and particle physics.

{\bf Acknowledgments}
The author thank the professors C. O. Escobar, M. M. Guzzo, O. L. G. Peres, P. C. de Holanda and R. da Rocha for useful discussions and FAPESP (PD 04/13770-0) for the financial support.

\end{document}